\documentclass[aps,pra,twocolumn,amssymb,amsmath,superscriptaddress,reprint]{revtex4-1}
\usepackage[colorlinks=true,urlcolor=blue,citecolor=blue,linkcolor=blue]{hyperref}
\usepackage{graphicx,amsmath,amssymb,amscd,epsfig}
\usepackage{color}
\usepackage{placeins}
\usepackage{float}

\usepackage[normalem]{ulem}
\usepackage{xcolor}
\definecolor{orange}{RGB}{255,127,0}

\newcommand{\deleted}[1] {\iffalse{#1}\fi}

\begin{document}
\title{Artificial Neural Network Approach to the Analytic Continuation Problem}

\author{Romain Fournier}
\affiliation{Institute of Physics, Ecole Polytechnique F\'ed\'erale de Lausanne (EPFL), CH-1015 Lausanne, Switzerland}

\author{Lei Wang}
\affiliation{
Institute of Physics, Chinese Academy of Sciences, Beijing 100190, China}

\author{Oleg V. Yazyev}
\email[E-mail: ]{oleg.yazyev@epfl.ch}
\affiliation{Institute of Physics, Ecole Polytechnique F\'ed\'erale de Lausanne (EPFL), CH-1015 Lausanne, Switzerland}
\affiliation{National Centre for Computational Design and Discovery of Novel Materials MARVEL, Ecole Polytechnique F\'{e}d\'{e}rale de Lausanne (EPFL), CH-1015 Lausanne, Switzerland}

\author{QuanSheng Wu}
\email[E-mail: ]{quansheng.wu@epfl.ch}
\affiliation{Institute of Physics, Ecole Polytechnique F\'ed\'erale de Lausanne (EPFL), CH-1015 Lausanne, Switzerland}
\affiliation{National Centre for Computational Design and Discovery of Novel Materials MARVEL, Ecole Polytechnique F\'{e}d\'{e}rale de Lausanne (EPFL), CH-1015 Lausanne, Switzerland}

\date{\today}

\begin{abstract}

Inverse problems are encountered in many domains of physics, with analytic continuation of the imaginary Green's function into the real frequency domain being a particularly important example.
However, the analytic continuation problem is ill-defined and currently no analytic transformation for solving it is known. We present a general framework for building an artificial neural network (ANN) that solves this task with a supervised learning approach.
Application of the ANN approach to the quantum Monte Carlo and simulated Green's function data demonstrates its high accuracy. 
By comparing with the commonly used maximum entropy approach, we show that our method can reach the same level of accuracy for low-noise input data, while performing significantly better when the noise strength increases.
The computational cost of the proposed neural network approach  is reduced by almost three orders of magnitude compared to the maximum entropy method.
\end{abstract}

\pacs{Valid PACS appear here}
\pacs{--------}

\maketitle



Numerical simulations are playing an extensive role in a growing number of scientific disciplines. Most commonly, numerical simulations approximate a process or a field on a discretized map, taking as input an equation describing the model as well as initial and boundary conditions. Problems falling under this definition are known as direct or forward problems. However, in a number of situations it is required to reconstruct an approximation of the input data or the model that generated it given the observable data. 
Such problem definitions are known as {\it inverse problems} and are mostly ill-posed~\cite{ramm_inverse_2006,kabanikhin_inverse_2011}. One particularly important example is the Fredholm integral equation of the first kind, which takes the following form:   
 \begin{equation}
 	g(t)=k\circ f := \int_a^b k(t,s)f(s)ds,
 	\label{eq:main_eq}
 \end{equation}
 where $g(t)$ is the available quantity, $f(s)$ is the quantity of interest and $k(t,s)$ is the kernel.
 Formally speaking, one is interested in operators $k$~$\circ$ that are ill-conditioned or degenerated. Therefore, the formal inversion $f=k^{-1}\circ g$ is not a stable operation. Noise affecting the data may lead to arbitrarily large errors in the solutions.

This work focuses on two particularly important examples of the analytic continuation problem in quantum physics, which take the same form as Eq.~\ref{eq:main_eq}. First, we will consider a prototypical problem of quantum harmonic oscillator linearly coupled to a bath, with the quantity of interest being the power spectrum $I(\omega)$ related to the imaginary-time correlation function $c(\tau)$ by a two-sided Laplace kernel 
 \begin{equation}
	c(\tau)=\int_{-\infty}^{\infty}\mathrm{e}^{-\omega\big(\tau-\frac{\beta}{2}\big)}I(\omega)\mathrm{d\omega}.
	\label{eq:oscillator_relation}
  \end{equation}  
Then, we will focus on the reconstruction of the single-particle spectral density function $A(\omega)$ from the single-particle fermionic Green's function $G(\tau).$ These quantities are related through the following relation: 
	\begin{equation}
    	G(\tau)=-\int \frac{e^{-\omega\tau}}{1+e^{-\hbar\omega\beta}}A(\omega)d\omega .
    	\label{eq:fermionic_relation}
	\end{equation}

There exist several techniques for performing the analytic continuation (see Refs.~\onlinecite{jarrell_bayesian_1996,sandvik_1998,beach_2004,syljuaasen_2008,beach_2004}) that regularize the problem by making use of prior knowledge and build an algorithm that converges towards a pseudo-solution~\cite{kabanikhin_2008}.  Among these methods, the so-called maximum entropy (MaxEnt) approach is the most commonly used. 
In this method, prior knowledge is added by specifying a default distribution $m(\omega)$ that corresponds to the expected results in the absence of data. The algorithm iteratively searches for a distribution $f(s)$ that maximizes the entropy with respect to $m(\omega)$, but at the same time generates, using Eq.~\ref{eq:fermionic_relation}, a function $g(t)$ close to the data at disposal. In other words, starting with a distribution $f^i$, one computes the standard mean squared deviation 
\begin{equation}
	\chi^2\big[f^i\big]=\sum_{m,n}\Big(g^i(\tau_m)-g(\tau_n)\Big)^2 \sqrt{C_{mn}^{-1}},
\end{equation}
and the entropy 
\begin{equation}
	S\big[f^i\big] = \sum_k \Delta\omega \Big(f^i(\omega_k)-m(\omega_k)-f^i(\omega_k)\mathrm{ln}\frac{f^i(\omega_k)}{m(\omega_k)}\Big),
\end{equation}
where $C$ is the correlation matrix of the available data $g$. Standard optimization procedures allow computing the distribution $\tilde{f}$ that maximalizes
\begin{equation}
	Q\big[f\big]=\alpha S\big[f\big]-\frac{\chi^2\big[f\big]}{2},
\end{equation}
where $\alpha$ is a parameter that weights the relative importance between the entropy and the error terms. There exist several methods for fixing $\alpha$, that often yield different results when applied in practice. 

The use of statistical methods for solving Fredholm integral equations has emerged recently. These methods include the regularization of the problem and the acceleration of existing methods through dimensionality reduction of the Green's functions~\cite{otsuki_sparse_2017, wu_acceleration_2013}. Dahm and Keller pointed out that a similar context of the Fredholm inverse integral equation of the second kind maps onto a reinforcement learning framework~\cite{dahm_learning_2017}. More recently, several works have shown that a machine learning approach is suitable for solving inverse problems~\cite{Li2018NETTSI,8103129,arsenault_projected_2017}. The main idea of these data-driven approaches is to distill the prior knowledge into simulated training datasets allowing a higher flexibility in the regularization of the dataset compared to the MaxEnt method. Arsenault {\it et al.}~\cite{arsenault_projected_2017} illustrated the utility of this approach for the fermionic spectral density function (Eq.~\ref{eq:fermionic_relation}). This work considered a database of spectral functions that resemble experimental data and calculated their corresponding Green's functions. A kernel ridge regression performed on these data yielded results comparable to those obtained using MaxEnt. 

In this Letter, we use a supervised learning approach for solving a physically more relevant scenario with known Hamiltonian and the data of interest obtained from quantum Monte Carlo (QMC) simulations. We use artificial neural networks (ANNs) as a convenient framework. The universal approximation theorem ensures that ANNs can approximate any kind of continuous functions under mild assumptions~\cite{Cybenko1989}. Furthermore, availability of powerful libraries allows for an efficient implementation of different ANN architectures that can take advantage of data structures, thus making ANNs a very versatile tool. We start by briefly describing how we generate the training dataset based on the physical parameters of the problem. We then present general training steps that ensure that the model has sufficient representative capacity and does not overfit the training data. The results obtained using the ANN and MaxEnt approaches applied to the QMC data are compared. Finally, we underline the versatility of our approach and its remarkable robustness against noisy input data by applying it to the fermionic kernel of~Eq.~\ref{eq:fermionic_relation} and comparing our results with a popular implementation of MaxEnt.

\begin{figure}[t!]
	\includegraphics{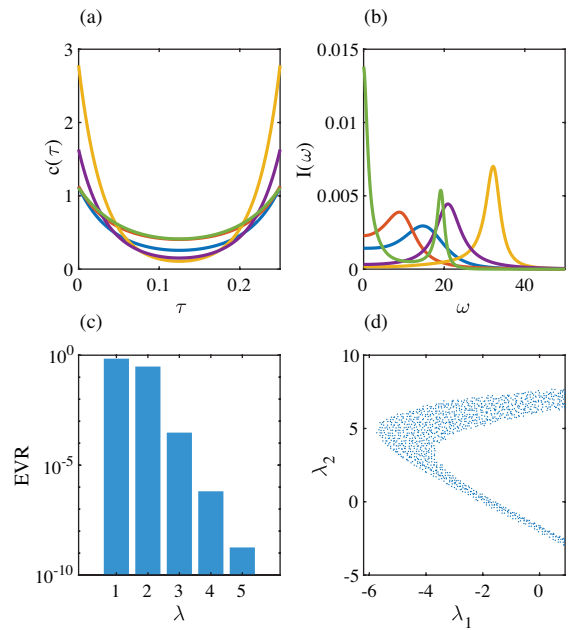} 
	\caption{(a) Examples of imaginary-time correlation functions present in the validation set and (b) their corresponding spectral density functions. (c) Illustration of the rapid decrease of the explained variance ratio (EVR) for the first five principal components. (d) Projection of data on the first two principal components.}
	\label{fig:database}
\end{figure}

To test our approach, we choose to study a system that has an analytical solution, yet proved to be difficult to solve using MaxEnt. We consider the time-correlation function of the position operator for a harmonic oscillator linearly coupled to an ideal environment. The Hamiltonian takes the following form:
 \begin{equation}
 	\hat{H}=\frac{\hat{p}^2 }{2m}+\hat{V}\left(\hat{x}\right)+\sum_{\alpha }\left\lbrace \frac{\hat{p}_{\alpha }^2 }{2m}+\frac{1}{2}m_{\alpha } {\omega_{\alpha } }^2 \hat{x}_{\alpha }^2 -{{\hat{x}c}}_{\alpha } \hat{x}_{\alpha }\right\rbrace,
 \end{equation}
where greek subscripts denote the bath variables, and $V(\hat{x})=\frac{1}{2}m {\omega_{0} }^2 \hat{x}^2$ is a harmonic potential of frequency $\omega_{0} $. The imaginary-time correlation function $ {c(\tau)=\big<\hat{x}(\tau)\hat{x}(0)\big>}$ can be computed within finite statistical error using QMC simulations, while the quantity of interest is the power spectrum $I(\omega)$ related to $c(\tau)$ through~Eq.~\ref{eq:oscillator_relation}. Following the work of Straub {\it et al.}~\cite{Straub1988}, we consider a bath with spectral density function 
  \begin{equation}
 	J_B(\omega)=\frac{1}{\pi} \int_0^{\infty } {dt}\;\mathrm{cos}\left(\omega t\right)\xi \left(t\right),
\end{equation}
where $\xi(t)$ is the classical friction kernel. For simulations of a fluid of Lennard-Jones particles, it takes the following form~\cite{Gallicchio1998}
\begin{equation}
	\xi \left(t\right)=\xi_0 \left\lbrace e^{-\alpha_1 {\left({ft}\right)}^2 } \left\lbrack 1+a_1 {\left({ft}\right)}^4 \right\rbrack +a_2 {\left({ft}\right)}^4 e^{-\alpha_2 {\left({ft}\right)}^2 } \right\rbrace.
	\label{eq:friction_kernel}
 	\end{equation}
This system has an analytic solution which relates Eq.~\ref{eq:friction_kernel} to the power spectrum \cite{SM},
providing an elegant way to generate physically relevant training data. A rich variety of spectra was generated from uniform distributions of the friction kernel parameters \cite{param1}, examples of which, as well as their corresponding imaginary-time correlation functions, are shown in Figs.~\ref{fig:database}a,b.

A simple approach to using machine learning for solving the inverse problem would be to take $\{c(\tau_i)\}$ as the input, and $\{A(\omega_i)\}$ as the output of our model. However, working in high-dimensional spaces leads to numerous problems commonly referred to as the {\it curse of dimensionality}. In absence of simplifying assumptions, the amount of data necessary to approximate a function to a given accuracy grows exponentially with the number of dimensions. This is due to the fact that for a given amount of data, the parameter space becomes more sparse when the dimensionality increases, which is also reflected in the {\it empty space phenomenon} term. Therefore, we will present a more compact representation of the input data to facilitate the learning process of the model. 
The simplest and most intuitive method of dimensionality reduction is the principal components analysis (PCA)~\cite{lee_nonlinear_2007}. The PCA relies on finding a linear transformation that defines a new orthogonal basis, whose components are aligned with the directions where the data displays the maximum variance. 
The fraction of variance covered by each principal component, the so-called explained variance ratio (EVR), provides a quantitative way of choosing the number of dimensions that need to be retained. The EVR of the training dataset shown in Fig.~\ref{fig:database}c decays exponentially fast with the component index, thus allowing to keep only the first three components with a loss of about $10^{-6}$ EVR.
Figure~\ref{fig:database}d shows the data projected on the first two components. The apparent structure underlines the limits of PCA, which can only suppress linear correlations between components. Non-linear methods such as t-SNE can be employed to further disentangle the data. 

\begin{figure}
	\includegraphics[width=8cm]{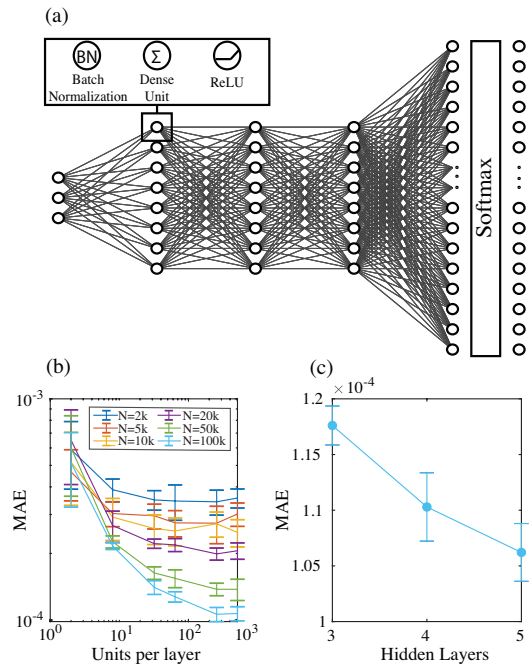}
	\caption{(a) Example of architecture with $m=3$ and $k=8$. The last layers have 1024 nodes. (b) Validation MAE at the end of the training phase for different training set sizes $N$ and units per layer $k$. (c) Same quantity for different number of hidden layers $(m+1)$ for $N=100000$ and $k=256$.}
	\label{fig:training}
\end{figure}

There exists no universal way to build a neural network, but prior knowledge of the problem can help designing it.
In this work, we consider architectures of $m$ fully connected feed-forward layers of $k$ units followed by one of 1024 units. The lack of structure in the three-dimensional input data suggests that the use of more complex architectures is not relevant. Batch normalizations between the layers and Rectifier Linear Units (ReLU), respectively, help the learning process and allow the model to find a non-linear relation between the input and the output. A final softmax output layer ensures the output to have the properties of a probability distribution. An example of an architecture with $m=3$ and $k=8$ is shown in Fig.~\ref{fig:training}a. The input data consist of the first three principal components of the correlation function, while the output dimension is 1024.

The training procedure was performed using the Adam optimizer~\cite{kingma_adam:_2014}, early stopping~\cite{goodfellow_deep_2016} and the mean absolute error (MAE) as the loss function
\begin{equation}
	 \mathrm{MAE}=\frac{1}{N}\sum_{i=1}^N|
	 I(\omega_i)-\hat{I}(\omega_i)|,
\end{equation} 
where $I(\omega_i)$ and $\hat{I}(\omega_i)$ are respectively the spectral density function present in the dataset and the one predicted by the model, both of them evaluated at $\omega_i$.
 This training method efficiently avoids overfitting issues. Performance of different architectures is assessed by averaging the results over ten instances after their training. The validation MAE values at the end of the training phase are shown in Fig.~\ref{fig:training}b for a fixed number of hidden layers ($m=4$) at different $k$ and number of training data entries $N$. Fig.~\ref{fig:training}c keeps $k=256$ and $N=100000$ constant and compares the results for different numbers of layers. As expected, increasing the number of parameters decreases the final validation MAE, until it reaches a plateau, whose value drops by adding data. We stopped training at $m=4$, $k=256$ and $N=100000$, achieving sufficient performance of this architecture.

We can now test our approach by returning to the initial problem. We computed the imaginary time correlation function $c(\tau)$ using QMC simulations. We set the parameters of the friction kernel (Eq.~\ref{eq:friction_kernel}) to $\xi_0=225$, $a_1=1.486\times10^5$, $a_2=285$, $\alpha_1=903$, $\alpha_2=75.0$ and $f=0.2$, and the bare oscillator frequency $\omega_0=20$. This system is known to be difficult to solve using MaxEnt due to its long relaxation time \cite{Krilov2001}. The imaginary time function $\hat{c}(\tau)$ was sampled on 64 slices $\tau_i \in [0,\beta]$, with $\beta=0.25$. This relatively large sampling induces errors due to the Trotter approximation, which makes it even harder to obtain the power spectrum \cite{Krilov2001}. Different slices were computed on different simulations to ensure their independence. The correlation function was computed from a total of $6\times10^6$ simulations divided into 300 blocks. Further details can be found in the Supplemental Material document~\cite{SM}. Before using ANN to obtain the imaginary-time correlation function, we must express the data on the same basis as the one used for the training. The QMC data were therefore transformed using the same PCA as the training data, interpolating the missing imaginary-time slices using cubic splines. 

\begin{figure}
	\includegraphics{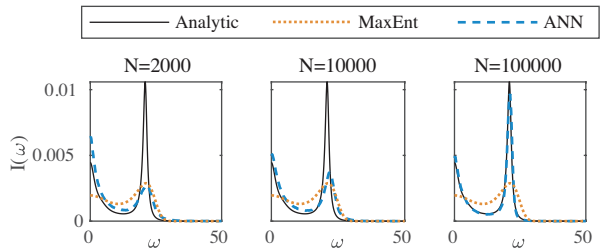}
	\caption{Analytic continuation of the QMC data performed using MaxEnt and the proposed ANN approach for different training set sizes $N$.}
	\label{fig:qmc}
\end{figure}

Figure~\ref{fig:qmc} shows that the essential features, the low-frequency decay and the peak at around $\omega=20$, are well reproduced by our ANN approach even for the model trained on 2000 data entries only. The model trained on the entire dataset of 100000 entries shows almost perfect agreement with the analytic solution. On the other hand, MaxEnt fails to provide accurate results. We would like to point out, however,  that better results using MaxEnt can be obtained by computing the correlation function $c(\tau)$ on a larger number of points \cite{Krilov2001}. 

To complete the validation of the proposed ANN model, we benchmark it using the methodology applied in the machine learning work of Arsenault {\it et al.} \cite{arsenault_projected_2017}. We seek to recover the electron single-particle spectral density in the real frequency domain $A(\omega)$ from the fermionic Green's function $G(\tau)$ in imaginary time domain, with the two quantities related through Eq.~\ref{eq:fermionic_relation}. The model spectral densities $A(\omega)$ are defined as a sum of uncorrelated Gaussian distributions with one peak constrained to be located close to the origin \cite{param2}.
Following previous works, we perform the dimensionality reduction by working in the orthogonal basis of Legendre polynomials and keeping first 64 coefficients \cite{huang_kernel_2016,wu_acceleration_2013}. Since the quantum Monte Carlo data is noisy by nature, we also generated test sets in which the Green's functions were corrupted by errors as 
$\tilde{G}(\tau_i)=G(\tau_i)+\epsilon_i$,
where $\epsilon_i$ are independent random variables normally distributed with standard deviation $\eta$.
Details of the dataset deneration and training can be found in the Supplemental Material~\cite{SM}. 

Fig.~\ref{fig:fermion}a provides a qualitative comparison of the results of our ANN model applied to unseen data and the MaxEnt implementation of Levy {\it et al.} \cite{levy_implementation_2017} for three different noise levels, $\eta=10^{-5}$, $10^{-3}$ and $10^{-1}$. 
The level of noise was provided as parameter for MaxEnt and used to select the network for ANN, as explained above. 
In these examples, both methods predict $A(\omega)$ accurately for the lowest level of noise. However, at $\eta=10^{-3}$ MaxEnt tends to suppress peaks in the predicted spectral function $\hat{A}(\omega)$, while in the case of our ANN model this tendency is much less pronounced. At the highest level of noise $\eta=10^{-1}$, our ANN model is able to reproduce most peaks, whereas MaxEnt flattens the distribution entirely (Fig.~\ref{fig:fermion}). 
Fig.~\ref{fig:fermion}b displays the MAE distributions of both the ANN and MaxEnt methods. It is clear that the mean MAE of the ANN model remains lower than the one of MaxEnt at all noise levels (Fig.~\ref{fig:fermion}c) and it shows a smaller spread. ANN outperfoms MaxEnt and behaves even better with increasing the noise level. 

\begin{figure*}
	\includegraphics[width=15.5cm]{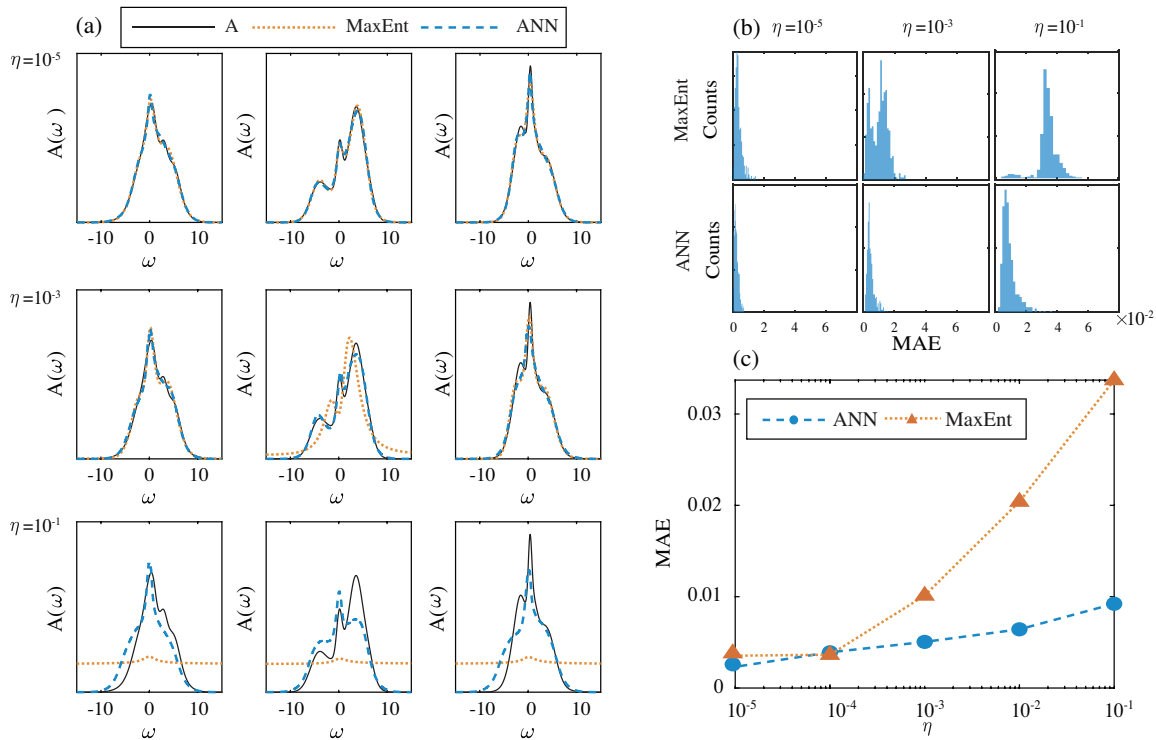}
	\caption{(a) Comparison of the starting spectral functions $A(\omega)$ (solid lines) with the predicted $\hat{A}(\omega)$ calculated using the MaxEnt approach and the proposed ANN model at different noise levels $\eta$ for three examples of spectral density functions not present in the training dataset. (b) MAE distributions for the two methods and different noise levels $\eta$. (c) Comparison of average MAE  for the two methods and different noise levels $\eta$. }
	\label{fig:fermion}
\end{figure*}

As a final note, we would like to underline the computational efficiency of our approach compared to MaxEnt. ANN allows a direct mapping between Green's functions and the spectral densities. In that sense, we can define it as a direct way of solving the problem. In contrast, MaxEnt is an iterative method which requires generating trial functions until convergence is reached. In our computational setup, the time required for converting 500 $(G(\tau),A(\omega))$ pairs with $\eta=10^{-5}$ is 5 seconds in the case of ANN (including the time to load libraries), while MaxEnt would need 51 minutes using the same setup.

To summarize, thanks to the stability of the forward problem, we have built an artificial neural network that solves the analytic continuation problem with an accuracy similar to that of the commonly used maximum entropy approach, but at a fraction of computational cost.  
We have also shown that our ANN model performs better for QMC data or noisy inputs. Adding more data and increasing the number of parameters can further improve the accuracy, although  training must be performed with care to avoid overfitting issues. The great representative capacity of deep neural networks suggests that other inverse problem can be solved in a similar way, provided that a dataset can be constructed. Such datasets may be derived using available experimental results combined with data augmentation techniques. The resulting models have the advantage of directly benefiting from prior knowledge and no longer relying on parameter tuning.  

Trained models resulting from our work can be obtained from public repository \cite{project_repository}. 

We thank Xi Dai for helpful discussions at the early stage of this project. O.V.Y. and Q.W. acknowledge support by NCCR Marvel. L.W. acknowledges support by the the National Natural Science Foundation of China under Grant No. 11774398. Computations were performed at the Swiss National Supercomputing Centre (CSCS) under project s832 and the facilities of Scientific IT and Application Support Center of EPFL.

\noindent
{\it Note:} Another work using convolutional neural networks for solving the analytical continuation problem~\cite{Yoon2018} appeared when the present manuscript was in preparation.



\begin{thebibliography}{25}
\bibitem [{param1()}]{param1}%
  \BibitemOpen
  \href@noop {} {}\bibinfo {note} {The friction kernel paramerters were sampled from the following intevals
$\xi_0 \in\lbrack100,300\rbrack$, $a_1 \in\lbrack13500,23500\rbrack$, $a_2 \in\lbrack150,300\rbrack$, $\alpha_1\in\lbrack800,1300\rbrack$, $\alpha_2 \in\lbrack70,100\rbrack$, $f\in\lbrack0.12,1.12\rbrack$.}
\BibitemShut {Stop}%
\bibitem [{param2()}]{param2}%
  \BibitemOpen
  \href@noop {} {}\bibinfo {note} {The model spectral densities $A(\omega)$ are defined as a sum of uncorrelated Gaussian distributions
\begin{equation}
    A_j(\omega)=\frac{1}{N_R}\sum_{i=0}^{R_j}\exp \left( -\frac{(\omega-\mu_i)^2 }{2\sigma_i^2} \right) ,
    \label{Aomega}
\end{equation}
where the frequencies $\omega \in [-\Omega_0,\Omega_0]$, the maximum frequency $\Omega_0=15$~eV, the centers of the peaks $\mu_j \in [-6,6]$~eV, the number of Gaussian distributions $R_j \in {1,2,...,21}$ and their broadening $\sigma_j \in [0.1,1]$~eV. The prefactor $1/N_R$ is used to normalize the spectrum. Parameters $R_j$, $\mu_j$ and $\sigma_j$ are uniformly sampled over the above-mentioned ranges, with the exception of $\mu_0$ that is constrained to be located close to the origin ($\mu_0 \in [-0.5, 0.5]$~eV).}\BibitemShut {Stop}%
\end{thebibliography}

%


\end{document}